\documentclass[iop,apjl]{emulateapj}

\usepackage{amsmath}
\usepackage{enumitem}
\usepackage{apjfonts}
\usepackage{hyperref}

\begin{document}

\title{First Interstellar HCO$^+$ Maser}
\shorttitle{First Interstellar HCO$^+$ Maser}
\author{Nicholas S. Hakobian}
\author{Richard M. Crutcher}
\shortauthors{Hakobian \& Crutcher}
\affil{University of Illinois at Urbana-Champaign, 1002 W. Green St. Urbana, IL 61801; \textit{nhakobi2@illinois.edu}, \textit{crutcher@illinois.edu}}

\journalinfo{\textrm{Accepted by ApJL, September 10, 2012}}
\submitted{Accepted September 10, 2012}

\begin{abstract}
  A previously unseen maser in the J = 1 - 0 transition of HCO$^+$ has
  been detected by the Combined Array for Millimeter-wave Astronomy
  (CARMA). A sub-arcsecond map was produced of the 2 arcmin$^2$ region
  around DR21(OH), which has had previous detections of OH and
  methanol masers. This new object has remained undetected until now
  due to its extremely compact size. The object has a brightness
  temperature of $>$ 2500 K and a FWHM linewidth of 0.497 km s$^{-1}$,
  both of which suggest non-thermal line emission consistent with an
  unsaturated maser. This object coincides in position and velocity
  with the methanol maser named DR21(OH)-1 by \citet{plambeck90}.  No
  compact HCO$^+$ emission was present in the CARMA data towards the
  other methanol masers described in that paper. These new results
  support the theory introduced in \citet{plambeck90} that these
  masers likely arise from strong outflows interacting with low mass,
  high density pockets of molecular gas. This is further supported by
  recent observations of a CO outflow by \citet{zapata12} that traces
  the outflow edges and confirms that the maser position lies along
  the edge of the outflow where interaction with molecular tracers can
  occur.
\end{abstract}


\maketitle

\section{Introduction}
  DR21(OH) is a part of a massive star forming region in the DR21
  molecular cloud complex. It lies along a ridge of strong molecular
  emission extending from DR21 in the south to W75N in the north. As
  with many objects of its type, DR21(OH) also contains numerous sites
  of strong maser emission in many species. Named for its particularly
  strong OH-masers, it was determined that the OH and H$_2$O masers
  were strongly coincident with millimeter continuum sources
  \citep{padin89}.

  DR21(OH) is particularly significant due to the detection of
  methanol masers over a wide range of frequencies. \citet{batrla88}
  detected methanol masers in the 81.6 GHz and 84 GHz methanol
  transitions. \citet{plambeck90} conducted a study of 95 GHz methanol
  emission and found four strong methanol masers across a region
  associated with the MM2 millimeter source. \citet{slysh97} presented
  detections of a previously unseen 133 GHz methanol maser. More
  recently, \citet{araya09} present an extensive survey of over 30 44
  GHz methanol masers, which appear to trace the shock fronts of two
  bow shocks along the red and blue shifted lobes of an outflow
  generated by MM2. \citet{fish11} present an analysis of all methanol
  maser detections and determine that all of the known masers appear
  to come from the interface region along the shock fronts of the
  outflow. The brightest of these masers appear along the western tip
  of the outflow and occur in a narrow velocity range of approximately
  0.3-0.5 km s$^{-1}$. Additionally, they determine that all the
  methanol maser transitions are Class I masers, which arise from
  collisional excitations and are considered to be caused by shocks,
  especially those that arise from outflows, as opposed to Class II
  masers, in which the pumping mechanism is primarily derived from
  external radiation. Recent submillimeter continuum and 1mm CO
  detections of the outflow morphology \citep{zapata12} give further
  support to the above model. In this letter we report the first
  detection of an HCO$^+$ maser in the interstellar medium.

\section{Observations}
\begin{figure*}[t]
    \begin{center}
    \includegraphics[angle=-90]{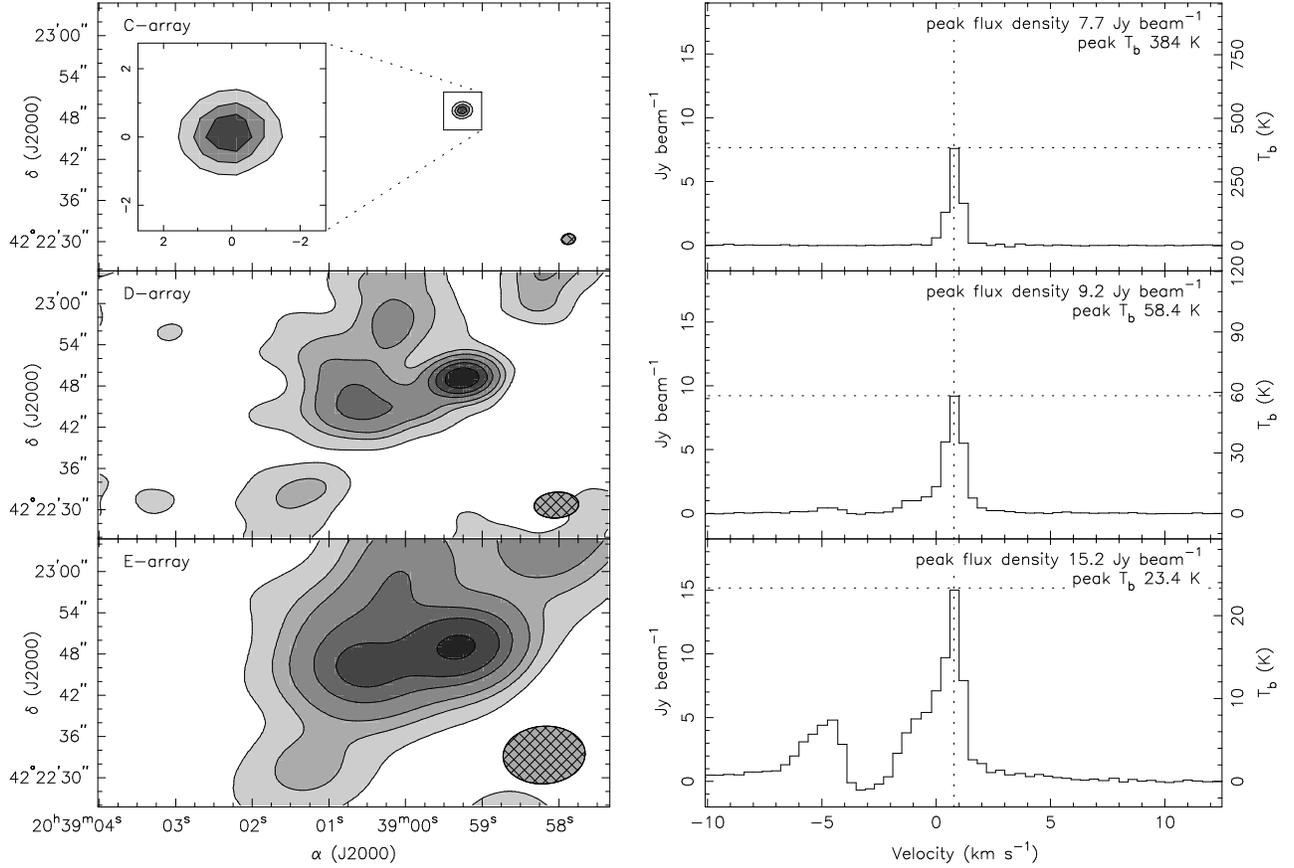}
    \caption{Comparison of maser emission in CARMA C, D, and
      E-arrays. Spectra (right) are of the peak positions in each map
      (left) of the 0.78 km s$^{-1}$ channel, visible in shaded
      contours. The y-axis of the spectra are shown in both units of
      Flux density (Jy beam$^{-1}$) and Brightness Temperature (K) to
      emphasize the effect of the decreasing source/beam size. The
      beam sizes are 2.0'' x 1.5'', 6.4'' x 3.7'', and 11.9'' x 8.4''
      for the C, D, and E-array maps, respectively. The contour levels
      for the C-array map are 0.25, 0.50, and 0.75 times the peak flux
      density. The contour levels for the D and E-array maps are 0.01,
      0.075, 0.15, 0.30, 0.45, and 0.7 times the peak flux
      density. The RMS noise of the spectra are 0.042, 0.028, and
      0.056 Jy beam$^{-1}$ for the C, D, and E array maps,
      respectively.\label{maser_a}}
    \end{center}
  \end{figure*}

\begin{figure*}[t]
    \begin{center}
    \includegraphics[angle=-90]{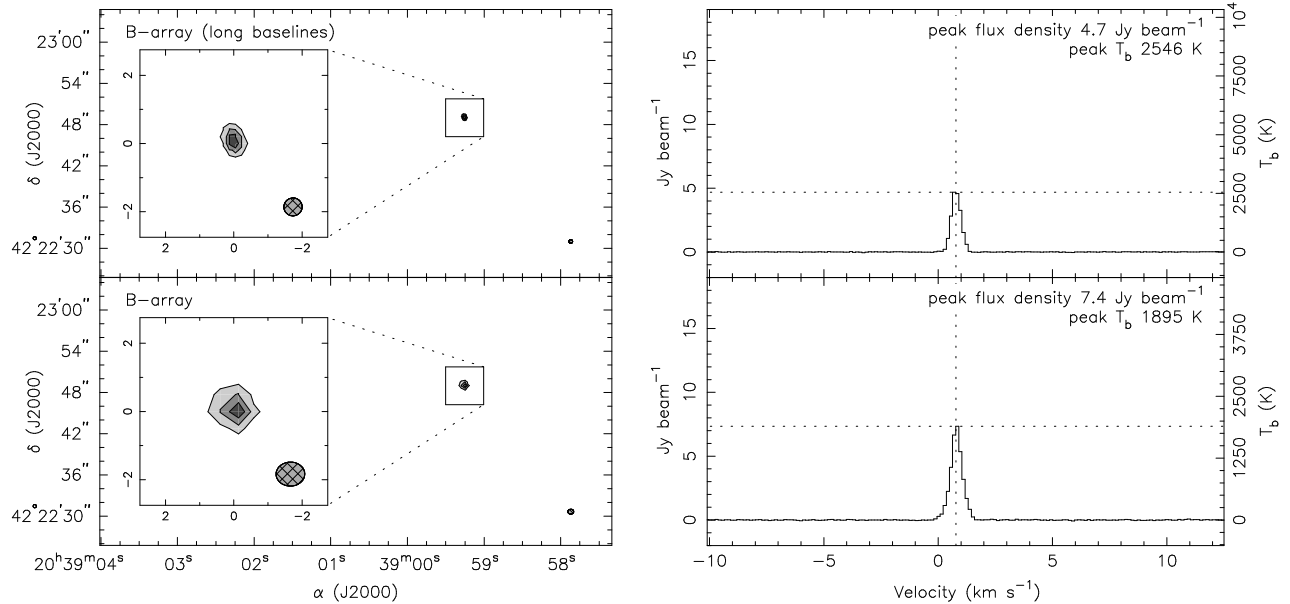}
    \caption{Comparison of maser emission in CARMA B-array. Spectra
      (right) are of the peak positions in each map (left) of the 0.78
      km s$^{-1}$ channel, visible in shaded contours. The y-axis of
      the spectra are shown in both units of Flux density (Jy
      beam$^{-1}$) and Brightness Temperature (K) to emphasize the
      effect of the decreasing source/beam size. The maps feature an
      inlay showing the 4 arcsecond$^2$ region around the maser due to
      its compact size. The top panel is of B-array data with
      baselines $>$ 140k$\lambda$. The beam sizes are 0.54'' x 0.52''
      for the top panel, and 0.85'' x 0.70'' for the bottom panel. The
      contour levels are 0.25, 0.50, and 0.75 times the peak flux
      density. The RMS noise of the spectra are 0.015 and 0.026 Jy
      beam$^{-1}$, for the top and bottom map,
      respectively.\label{maser_b}}
    \end{center}
  \end{figure*}

  DR21(OH) was observed as part of an ongoing survey of sources
  \citep[see][]{hakobian11} with strong magnetic field detections
  \citep{falgarone08,crutcher99}. A 2 square arcminute region around
  DR21(OH) was mapped with observations by CARMA \citep{bock06} in the
  C, D, and E array configurations with a combined effective beam size
  of 6.3" x 4.0", and a spectral resolution of 0.41 km s$^{-1}$. This
  was achieved by observing in the 8 MHz bandwidth correlator mode
  with 63 channels. The J = 1-0 transition of HCO$^+$, with a rest
  frequency of 89.188526 GHz, was simultaneously observed with HCN,
  N$_2$H$^+$, and two continuum bands. A bright, unresolved, and
  unusually narrow line feature was detected in the map of HCO$^+$
  (Fig. \ref{maser_a}), at
  20$^{\textrm{h}}$38$^{\textrm{m}}$59.3$^{\textrm{s}}$
  +42$^{\textrm{o}}$22'49.0'' (J2000) with a peak at
  V$_{\textrm{LSR}}=$ 0.78 km s$^{-1}$. Each track was inspected to
  rule out the possibility of a transient instrumental issue; however,
  the compact source was visible in each track. The spatial and
  spectral location of the source additionally does not change between
  each of the tracks, indicating that an instrumental issue is
  unlikely. Follow-up observations in CARMA B-array configuration
  (Fig. \ref{maser_b}) were performed in December 2011 in order to
  further constrain the physical size and brightness of the source. In
  total, five tracks of data were observed for a total of 18.75 hours:
  4 hours each in B, C, and E arrays, and 6.75 hours in D array (over
  two tracks). Data were calibrated and imaged with the CARMA MIRIAD
  software package. The phase calibrator used in all tracks was
  2038+513. Due to the strength of the calibrator, we were able to use
  the standard phase calibration techniques. The B-array track has an
  improved spectral resolution of 0.14 km s$^{-1}$ over the other
  tracks, due to being observed with the expanded CARMA 8-band
  correlator in the 8 MHz bandwidth mode with 384 Hanning smoothed
  spectral channels (Figure \ref{maser_b} shows the 192 spectrally
  independent channels).

\section{Data / Analysis}

  \begin{figure*}[t]
    \begin{center}
    \includegraphics[angle=-90, width=\linewidth]{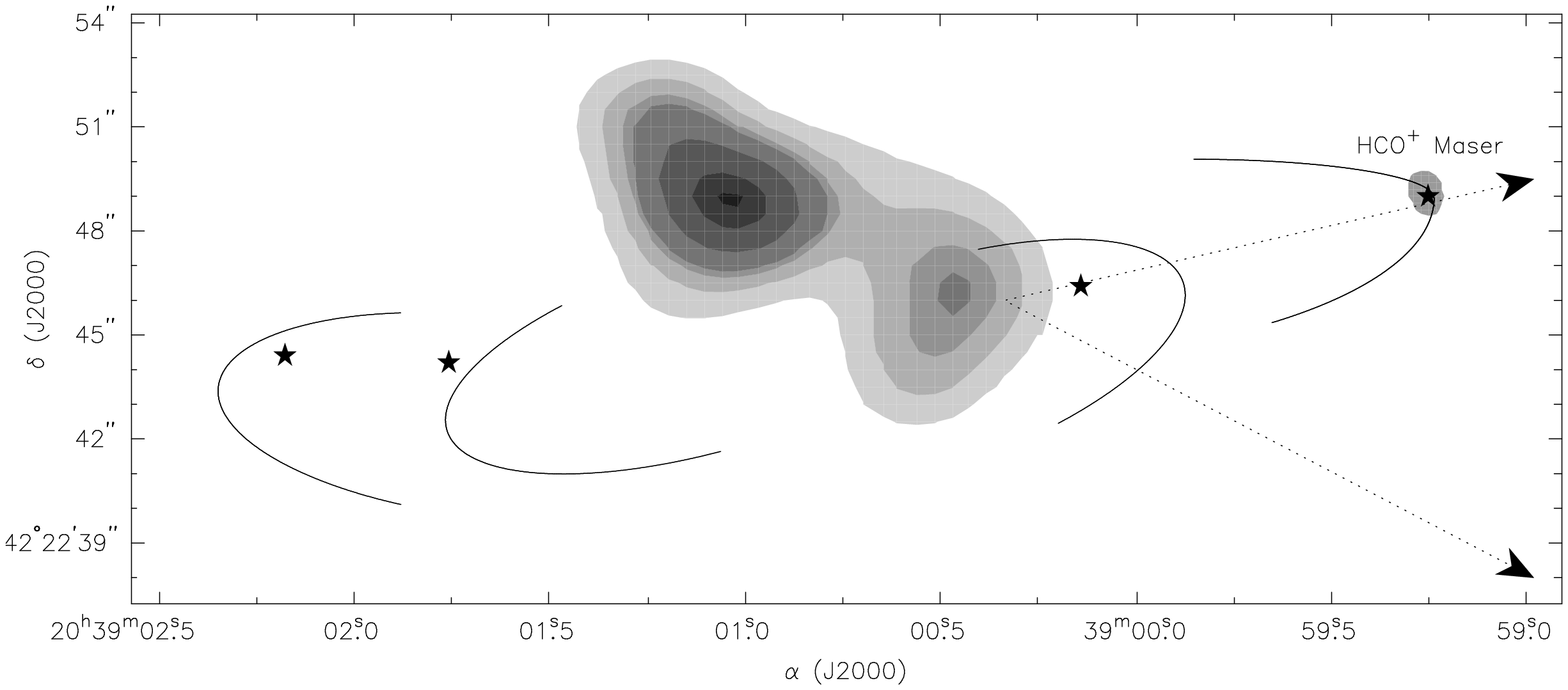}
    \caption{Diagram of the position of the HCO$^+$ maser with respect
      to other features of DR21(OH). The shaded contours are of the
      112 GHz continuum of DR21(OH). The dotted arrows represent the
      CO outflow as observed by \citet{zapata12}. The shaded oval
      represents the extent of the HCO$^+$ object as reported in this
      paper. The stars represent the positions of methanol masers from
      \citet{plambeck90}. The black curves represent the two sets of
      bow shocks determined by the loci of 44 GHz methanol masers from
      \citet{araya09}. The positions of the masers along the edge of
      the outflow give strong support to the theory that shocked
      interactions between outflows and small, high density clumps of
      molecular gas give rise to these objects. The contour levels of
      the continuum emission are 0.3, 0.4, 0.49, 0.54, 0.7, 0.9, and
      0.99 times the peak flux density of 0.103 Jy beam$^{-1}$ in
      order to show the peaks of both continuum
      sources.\label{diagram}}
    \end{center}
  \end{figure*}

Using the high resolution B-array data, we find that the angular size
of this object is less than $\sim$ 0.8 arcseconds and that it has a
brightness temperature $>$ 1900 K. Since the source remained
unresolved in B-array, the parameters of this object can be further
constrained by using only long baseline components of the data
set. This effectively ``resolves out'' larger scale structure, leaving
a constrained map behind. To achieve this, we limited the map to use
baselines larger than 140 k$\lambda$ ($\sim$ 470 meters). This
procedure resulted in a map with an effective beam size of
$\sim$0.54'' x 0.52'' and a brightness temperature of $>$ 2500 K. The
source appears to be unresolved even at this resolution (Fig
\ref{maser_b}, top panel).

If the source were dominated by thermal emission, a lower bound on its
linewidth can be estimated by calculating the degree by which the
emission would be thermally broadened. The calculated full-width at
half maximum (FWHM) linewidth would then be purely a function of the
effective brightness temperature:
\begin{equation}
  \begin{aligned}
    \Delta f &= \sqrt{\frac{8 k T \ln 2}{m c^2}}f_0, \;\;\;\;
    \frac{\Delta f}{f_0} = \frac{\Delta v}{c}, \;\;\;\;
    m = N m_p \\
    \Delta v &= \sqrt{\frac{8 \ln 2 k}{m_p}} \; \sqrt{\frac{T}{N}} \\
             &= 0.2139 \; \sqrt{\frac{T}{N}} \;\; \textrm{km s}^{-1}
  \end{aligned}
\end{equation}
where $T$ is the brightness temperature of the source and $N$ is the
atomic weight of the molecule. The factor of $2\sqrt{2 \ln 2}$ comes
from the conversion of a Gaussian standard deviation to FWHM. For a
source with a brightness temperature of 2546 K and RMS noise of 8.6 K
(0.015 Jy beam$^{-1}$), the thermal linewidth would be 2.077 $\pm$
0.004 km s$^{-1}$, which is significantly greater than the observed
linewidth of 0.497 $\pm$ 0.002 km s$^{-1}$ (obtained by least-squares
gaussian fitting to the line profile). The maser linewidth corrected
for the instrumental smoothing is 0.477 $\pm$ 0.002 km s$^{-1}$. The
observed high brightness temperature and linewidth narrowing are good
indicators that this object is dominated by non-thermal emission. An
assumption as to the type of maser emission can be made from this
linewidth analysis. Saturated masers can have linewidths up to the
thermal linewidth, while unsaturated masers will have a linewidth
narrower than the thermal linewidth by a factor of 4 to 5
\citep{reid88}. Since our object has a linewidth that is 4.2 times
narrower than the thermal linewidth, it is consistent with an
unsaturated maser.

\citet{plambeck90} observed this region with BIMA at the 95 GHz
methanol line and discovered four methanol masers connected by large
scale methanol emission. The brightest of these four sources,
DR21(OH)-1, is centered at
20$^{\textrm{h}}$38$^{\textrm{m}}$59.24$^{\textrm{s}}$
$+$42$^{\textrm{o}}$22'49.04'' (J2000) with a peak at
V$_{\textrm{LSR}}$= 0.32 km s$^{-1}$, which is approximately 0.1''
from the center of our measured HCO$^+$ peak. The HCO$^+$ position has
an approximate positional error of 0.2'', and the methanol positions
have an error of 0.3''. From these data they estimated their methanol
maser had an angular diameter $<$4.4'' and a brightness temperature $>$
760 K. Three other methanol masers were also observed; however, there
is no evidence of companion HCO$^+$ masers (Fig. \ref{diagram}). This
could possibly be due to their relatively weaker strength or due to
the fact that the conditions necessary to produce an HCO$^+$ maser do
not exist at these other positions. The co-location of the HCO$^+$
object and the methanol maser suggest that both arise from similar
conditions. \citet{plambeck90} hypothesized that such regions could be
created from the interaction of an outflow and small clumps of dense
molecular gas.

In comparing the B-array maps (Figure \ref{maser_b}), it appears as if
the maser could be partially resolved since its intensity drops from
7.4 Jy beam$^{-1}$ to 4.7 Jy beam$^{-1}$ between them, with slight N-S
structure. \citet{araya09} reports that there are three methanol
masers within 2'' of the HCO$^+$ emission; one strong and two weaker
masers, slightly north and south of the strongest maser. If the
HCO$^+$ maser is also a complex of three individual sources oriented
in a N-S direction, the highest resolution B-array map may be
beginning to distinguish them, giving an appearance of being
resolved. In the D and E arrays, the sharp maser ``peak'' extends
$\sim$ 7.5 Jy beam$^{-1}$ above the extended emission, in agreement
with the flux density in the B-array spectrum.

A small amount of HCO$^+$ absorption is visible in the E-array map;
however, it is not apparent that the maser linewidth is affected by
self-absorbtion. In this map, the shortest baselines will sample large
enough size scales to include the western extension of the continuum
that peaks with DR21(OH)-MM2. Furthermore, the HCO$^+$ emission
includes two velocity components associated with extended gas, one at
-4.7 km s$^{-1}$, and another peaking at $\sim$ 0 km s$^{-1}$. The
peak positions of these two velocity components correspond to the
continuum peaks of MM1 and MM2 respectively. The high velocity wing of
the -4.7 km s$^{-1}$ component is affected by absorption by the
continuum in the E-array map, however, the 0 km s$^{-1}$ component is
not. The D and C array maps with smaller beam sizes (longer baselines)
are not affected by this continuum contamination or the extended
HCO$^+$ emission which is resolved out.

\citet{zapata12} performed 1mm observations of several spectral lines
around DR21(OH). Included with these observations are CO(2-1)
observations which trace an outflow from DR21(OH)-MM2. This outflow
appears to be in the plane of the sky (Fig. \ref{diagram}). From this
figure, we can see that both the methanol maser and our HCO$^+$ object
appear along the edge of the outflow. This region would be highly
shocked, and the energy released from this interaction has the
potential to power the maser.

\citet{goldsmith72} analyzed the J = 1 - 0 transition of CO and
determined that a large range of rotational excitation temperatures,
including population inversion, can be produced through collisional
excitation in the range of kinetic temperatures and densities found in
molecular clouds. It was also concluded that other linear molecules
with simple rotational structure, such as HCO$^+$, would have the same
result.  Using the RADEX radiative transfer package
\citep{vandertak07}, we performed a test calculation that showed that
for collisional interaction such as is suggested here in DR21(OH), it
is possible to achieve population inversion in the HCO$^+$ J = 1 - 0
transition with gas densities of $\sim 5 \times 10^5$ cm$^{-3}$ and
kinetic temperatures $\sim$50 K (consistent with kinetic temperatures
in outflow shocked regions). This result further supports the
conclusion that this HCO$^+$ is a maser.

\section{Conclusion}

Observations of DR21(OH) have revealed the presence of a compact
object which is dominated by non-thermal emission. The extremely
compact size of this object coupled with its large brightness leads to
the following conclusions: 
\begin{itemize}[noitemsep,nolistsep]
  \item The source is co-located with a known strong methanol maser.
  \item It lies along the edge of an outflow which gives support to
    previous theories that masers can arise due to the interactions of
    high velocity outflows with cold, dense clumps of molecular gas.
  \item Its small spatial size most likely prevented its detection
    before now; emission from this source would be beam diluted to
    levels indistinguishable from thermal emission.
  \item This object is very likely an unsaturated maser, the first
    observed in HCO$^+$.
\end{itemize}
In order to further confirm that this source is indeed a maser, future
observations to look for anomalous level populations in higher order
HCO$^+$ transitions should be performed.

\section{Acknowledgments}

Support for CARMA construction was derived from the states of
California, Illinois, and Maryland, the James S. McDonnell Foundation,
the Gordon and Betty Moore Foundation, the Kenneth T. and Eileen
L. Norris Foundation, the University of Chicago, the Associates of the
California Institute of Technology, and the National Science
Foundation. Ongoing CARMA development and operations are supported by
the National Science Foundation under a cooperative agreement (NSF AST
08-38226), and by the CARMA partner universities.
\ \\
\ \\
\ \\


\vfill

\begin{thebibliography}{}
\bibitem[Araya et al.(2009)]{araya09}Araya, E.D., Kurtz, S., Hofner, P., Linz, H., 2009, ApJ, 698, 1321
\bibitem[Batrla \& Menten(1988)]{batrla88}Batrla, W., and Menten, K.M., 1988, ApJ, 329, 117
\bibitem[Bock et al.(2006)]{bock06}Bock, D.C.-J., Bolatto, A.D., Hawkins, D.W., Kemball, A.J., Lamb, J.W., Plambeck, R.L., Pound, M.W., Scott, S.L., Woody, D.P., \& Wright, M.C.H., ``First results from CARMA: the combined array for research in millimeter-wave astronomy'', 2006, Proc. SPIE, 6267, 13
\bibitem[Crutcher(1999)]{crutcher99}Crutcher, R.M., 1999, ApJ, 520, 706
\bibitem[Falgarone et al.(2008)]{falgarone08}Falgarone, E., Troland, T. H., Crutcher, R. M., Paubert, G., 2008, A\&A, 487, 247
\bibitem[Fish et al.(2011)]{fish11}Fish, V.L., Muehlbrad, T.C., Sjouwerman, L.O., Strelnitski, V., Pihlstr\"{o}m, Y.M., Bourke, T.L., 2011, ApJ, 729, 14
\bibitem[Goldsmith(1972)]{goldsmith72}Goldsmith, P., 1972, ApJ, 176, 597
\bibitem[Hakobian \& Crutcher(2011)]{hakobian11}Hakobian, N.S., and Crutcher, R.M., 2011, ApJ, 733, 6
\bibitem[Padin et al.(1989)]{padin89}Padin, S., Sargent, A.I., Mundy, L.G., Scoville, N.Z., Woody, D.P., Leighton, R.B., Masson, C.R., Scott, S.L., Seling, T.V., Stapelfeldt, K.R., Terebey, S., 1989, ApJ, 337, 45
\bibitem[Plambeck \& Menten(1990)]{plambeck90}Plambeck, R.L., and Menten, K.M., 1990, ApJ, 364, 555
\bibitem[Reid \& Moran(1988)]{reid88}Reid, M.J., and Moran, J.M., 1988, ``Galactic and Extragalactic Radio Astronomy'', Editors: Verschuur, G.L., and Kellerman, K.I., p. 255-293
\bibitem[Slysh et al.(1997)]{slysh97}Slysh, V.I., Kalenskii, S.V., Val'tts, I.E., Golubev, V.V., 1997, ApJ, 478, 37
\bibitem[Van der Tak et al.(2007)]{vandertak07}Van der Tak, F.F.S., Black, J.H., Sch\"{o}ier, F.L., Jansen, D.J., van Dishoeck, E.F., 2007, A\&A, 467, 627
\bibitem[Zapata et al.(2012)]{zapata12}Zapata, L.A., Loinard, L., Su, Y.-N., Rodriguez, L.F., Menten, K.M., Patel, N., Galvan-Madrid, R., 2012, ApJ, 744, 86
\end{thebibliography}
\end{document}